\documentclass[]{elsarticle}
\usepackage{amsfonts}
\usepackage{amsmath}
\usepackage{amssymb}
\usepackage{txfonts}
\usepackage{pxfonts}
\usepackage{graphicx,bm,units,yfonts,yhmath}
\usepackage[table]{xcolor}
\usepackage{hyperref}

\newcommand{\I}{{\rm i}}

\begin{document}

\title{Mapping the Current-Current Correlation Function Near a Quantum Critical Point}
\author{Emil Prodan$^{1}$ and Jean Bellissard$^{2}$}
\address{\ $^{1}$ Department of Physics, Yeshiva University, New York, New York 10016, USA
\\$^2$ Department of Mathematics, Georgia Institute of Technology, Atlanta, Georgia, USA
}

\begin{abstract}
The current-current correlation function is a useful concept in the theory of electron transport in homogeneous solids. The finite-temperature conductivity tensor as well as Anderson's localization length can be computed entirely from this correlation function. Based on the critical behavior of these two physical quantities near the plateau-insulator or plateau-plateau transitions in the integer quantum Hall effect, we derive an asymptotic formula for the current-current correlation function, which enables us to make several theoretical predictions about its generic behavior. For the disordered Hofstadter model, we employ numerical simulations to map the current-current correlation function, obtain its asymptotic form near a critical point and confirm the theoretical predictions.    
\end{abstract}


\maketitle

\section{Introduction}

The quantum critical regime at the plateau-plateau transition \cite{WTP,KW,Pru1,KHKP1,KHKP2,DSMM,KHKP3,Huc,LVX} and at the plateau-insulator transition \cite{SchaijkPRL2000vc,DunfordaPE2000gf,PonomarenkoPE2000yt,PruiskenSSC2006tr,VisserJPCS2006cu,LangPRB2007yg} in the phase diagram of the integer quantum Hall effect (IQHE) has been the subject of intense experimental and theoretical scrutiny. As a result, the critical behaviors of the transport coefficients have been mapped with extraordinary experimental precision and accurate quantitative analyses of the scaling laws became possible. The results of these activities can be summarized as follows. First, the following scaling law of the resistivity tensor $\rho$ with temperature $T$ has been observed in most if not all the experiments:
\begin{equation}\label{TScaling}
\boxed{
\rho(E_F,T) =F\left( \big (E_F-E_c \big )\left(\frac{T}{T_0} \right )^{-\kappa}\right).
}
\end{equation}
A similar scaling law applies to the conductivity tensor $\sigma(E_F,T)$, which is just the inverse of $\rho$. In Eq.~\eqref{TScaling}, $\kappa$ is the finite-temperature scaling exponent, $F$ is a system-specific ({\it i.e.} non-universal) function, $T_0$ is a reference temperature and $E_c$ is the critical Fermi energy. For the plateau-insulator transition, the exponent seems to converge to $\kappa \approx 0.57$ \cite{LangPRB2007yg}, a value which is substantially larger than $\kappa = 0.42$ consistently reported for the plateau-plateau transition (see {\it e.g.} Ref.~\cite{LVX}). Understanding this difference is still an open problem. The finite-temperature scaling exponent can be related to the finite-size scaling exponent $\nu$ (see below), the latter being set by the asymptotic behavior of the Anderson localization length \cite{AAL} near the critical energy:
\begin{equation}\label{EScaling}
\boxed{
\Lambda(E_F) \sim \frac{\Lambda_0}{(E_F-E_c)^\nu }.
}
\end{equation}
The latest numerical estimates give $\nu = 2.58 \pm 0.03$ \cite{SO1,KMO,OSF,FHA,DET,AMS,SO2,OGE}, a value which is considerably larger than $\nu = 2.4$ reported by older simulations (see {\it e.g.} \cite{Huc}) and by the experimental papers, {\it e.g.} Ref.~\cite{LVX}. As one can see, the IQHE continue to surprise us and remains a subject of great interest.

\vspace{0.2cm}

As explained in Ref.~\cite{Pru1}, a relation between $\kappa$ and $\nu$ can be established as follows. At finite but low enough temperature, the thermal broadening of the Fermi-Dirac distribution does not play a significant role and the temperature-dependence of the transport coefficients is determined entirely by the dissipative processes, whose strength typically scales as $ T^{p}$ at low temperatures. The exponent $p$ is often referred to as the dynamical exponent for dissipation. Now, the dissipation determines a characteristic length, the Thouless length \cite{Tho}:
\begin{equation}
L_{\rm Th}(T) \sim T^{-p/2},
\end{equation} 
which can be interpreted as a temperature-induced finite effective size for the otherwise infinite system. Then the single-parameter scaling hypothesis, known to be strictly obeyed near the quantum transitions in IQHE \cite{Pru2}, infers that the transport coefficients are actually a function of the ratio:  
\begin{equation}\label{X1}
\chi =L_{\rm Th}(T)/\Lambda(E_F) \sim \Big ( (E_F-E_c) T^{-p/2\nu}\Big )^\nu.
\end{equation} 
For example, in this quantum critical regime, a sample will switch between a metallic and an insulating behavior at $\chi \approx 1$ when changing $E_F$ and $T$, but the switch happens in a manner which depends entirely on $\chi$ (see Ref.~\cite{Pru1} for more details). Comparing Eq.~\eqref{X1} with the argument of the function $F$ in \eqref{TScaling}, the following relation emerges:
\begin{equation}\label{Kappa}
\boxed{
\kappa=\frac{p}{2\nu}.
}
\end{equation}

Recently, one of the authors was involved with a numerical simulation \cite{SP} which reproduced qualitatively and quantitatively many of the conclusions drawn in \cite{LangPRB2007yg} via a careful analysis of the available experimental data on the plateau-insulator transition. The setting of the simulation was that of homogenous 2-dimensional disordered lattice systems under a uniform magnetic field and the framework was that of the transport theory for electrons developed in Refs.~\cite{BES,SBB1,SBB2}, which incorporates the dissipation effects. As we have seen above, this is absolutely necessary in order to understand the quantum critical regime. More specifically, the simulations reported in \cite{SP} were based on the finite-temperature Kubo-formula written in Eq.~\eqref{Kubo1} and simplified to the so called relaxation time approximation. Encouraged by the agreement between the simulation and the experiment, we employ here similar numerical techniques to further explore the physics of the plateau-insulator transition in IQHE.

\vspace{0.2cm}

This time, we switch our attention to the equivalent version of the Kubo-formula written in Eq.~\eqref{Kubo2}. There, the reader will notice the current-current correlation measure ${\rm d} m_{ij}(E,E')$, which over the year proved to be an extremely useful concept in the theory of electron transport \cite{LGP,KP}. As it is well known \cite{BES}, not only the transport coefficients but also Anderson's localization length can be expressed in terms of this correlation measure (see Eq.~\ref{Lambda}). For homogeneous systems, this measure can be defined as the infinite-volume limit \cite{Pas,HL,CGH}:
\begin{equation}\label{CCM1}
{\rm e}^2\int\limits _{\Delta E \times \Delta E'} {\rm d} m_{ij}(E,E')= \lim_{V \rightarrow \infty} \frac{1}{V} \sum_{(\epsilon_n,\epsilon_m) \in \Delta E \times \Delta E'} \langle \psi_n|J_i|\psi_m \rangle \langle \psi_m | J_j | \psi_n \rangle,
\end{equation}
where $(\epsilon_n, \psi_n)$ is the eigen-system of the finite-volume Hamiltonian and $\bm J$ is the current operator. Throughout, ${\rm e}$ will denote the charge of the electron. An alternative definition of the current-current correlation measure, without involving any the thermodynamic limit,  will be introduced in Section~3 after some background is developed. The current-current correlation function $f_{ij}(E,E')$ is defined as the density of the correlation measure:
\begin{equation}
{\rm d} m_{ij}(E,E')=f_{ij}(E,E')dE dE',
\end{equation}
and here we will be only dealing with the isotropic part $f = \frac{1}{2}\sum_{i} f_{ii}$. The fundamental assumption of our work is that such density exists and is continuous away from the critical point. This assumption seems reasonable since the energy spectrum is known from experiment to be dynamically localized away from the critical energy. In Ref.~\cite{BH}, for example, under certain assumptions on the disorder configuration space, the existence and continuity (even analyticity) of $f(E,E')$ was established for the Anderson model in a domain away from the diagonal $E=E'$. On the other hand, the existence of the diagonal $f(E,E)$ for random Schr\"odinger operators when $E$ in a region of dynamically localized spectrum was established in \cite{CGH} and, furthermore, it was shown there that $f(E,E)=0$ in such conditions. The existence and continuity of $f(E,E)$ can also be established \cite{Bellissard3} for the entire energy plane for the Gaussian random  Hamiltonians which behave as random matrices from the Gaussian ensembles and the Hamiltonians and the position operator are free random variables.

\vspace{0.2cm}

Under the assumption mentioned above, we derive the following asymptotic form of the current-current correlation function:
\begin{equation}\label{CCCF}
f(E,E')=g\left( \frac{E+E'-2E_c}{(E-E')^{\kappa/p}} \right ), \quad E,E' \simeq E_c
\end{equation}
which reproduces the critical behavior of the transport coefficients and of the Anderson localization length simultaneously, provided $\kappa$, $\nu$ and $p$ obey the constraint of Eq.~\eqref{Kappa}. Furthermore, based on several universal features of the transport coefficients, seen in the majority of the existing experimental data on IQHE, we put forward several predictions about the generic behavior of $f(E,E')$. In the second part of our work, we obtain a numerical representation of the current-current correlation function for the Hofstadter model \cite{Hof} with on-site disorder, around the critical energy of the plateau-insulator transition analyzed in Ref.~\cite{SP}. The numerically computed $f$ conforms with the theoretical predictions mentioned above. Furthermore, the function $g$ in Eq.~\eqref{CCCF} is found to be reasonably well represented by a Gaussian function.

\vspace{0.2cm}

We want to mention that, although our application deals exclusively with the metal-insulator transition in IQHE, both theoretical and numerical techniques are quite general. As such, the presentation is kept as broad as possible. Section 2 gives a brief overview of the class of homogeneous solid state systems for arbitrary space-dimension. The current-current correlation measure is introduced in Section 3 in this general setting. The scaling analysis is performed in Section 4 and it assumes only Eqs.~\eqref{TScaling} and \eqref{EScaling}. The principles of our numerical alghorithms are presented in Section 5 and the concrete numerical application to the disordered Hofstadter model is presented in Section 6.

\section{Homogeneous solid state systems}

Imagine that we have a stack of mesoscopic samples, all cut out from one big piece of a homogeneous material, such as a batch of copper prepared in a big furnace. It is an inescapable fact that, no matter how careful the fabrication process was, the samples differ at the microscopic scale. Yet, the macroscopic measurements performed on different samples will return identical values of the thermodynamic coefficients, within the accuracy of the instruments and of the measuring methods.  Explaining this empirical observation in simple terms is one goal of the present section (for more formal treatments see \cite{Bellissard1,Bellissard2}). In the process, we develop the background required in the following Sections and introduce the disordered model used in the numerical applications.

Solid state systems can be accurately and efficiently described by models defined on the lattice $\mathbb Z^d$. Here, $d$ is the effective dimension of the solid, which can be, for example, $d=1$ for polymer chains, carbon nanotubes, etc., $d=2$ for graphene, silicene, germanene, etc., and $d=3$ for the ordinary crystalline solids. Note that the actual physical system doesn't need to be perfectly straight or perfectly flat to be considered a lower dimensional atomic structure, because here the lattice $\mathbb Z^d$ serves primarily a labeling purpose rather than a physical representation. A node $\bm n$ of the lattice labels a unit cell of the solid, which typically contains more than one atom. Let the physical Bravais lattice of the crystal be generated by $\{\bm e_1,\ldots,\bm e_d\}$, which are vectors from $\mathbb R^d$. The chemically and physically relevant molecular orbitals affiliated with the unit cell $\bm n$ are represented by the vectors $|\bm n, \alpha \rangle $, $\alpha = 1,\ldots, N$, whose linear spann generates the Hilbert space of the model $\mathcal H = \mathbb C^N \otimes \ell^2(\mathbb Z^d)$. These orbitals can be rigorously  defined within the framework of maximally localized Wannier functions \cite{MV,BPC} and the Hamiltonian of a lattice model system can be derived empirically or from first principles. For the first method, by increasing the number $N$ of orbitals, the lattice models can be finely tuned on the available experimental data and subsequently be used to make predictions. The principles and the accuracy of the second method are described in great details in Ref.~\cite{MMY}.  

\vspace{0.2cm}

In the ideal case of strictly periodic solids, the lattice Hamiltonians take the generic form:
\begin{eqnarray}\label{PModel}
H_{\bm 0}= \sum_{{\bm n},\alpha; \bm m, \beta}  t_{{\bm n}-{\bm m}}^{\alpha \beta} \; |{\bm n},\alpha \rangle  \langle {\bm m},\beta | ,
\end{eqnarray}
where the sum typically runs only over a few neighboring unit cells and $t_{{\bm n}-{\bm m}}^{\alpha \beta}$ are just $c$-numbers. Note the dependence of the hopping amplitudes on the difference ${\bm n}-{\bm m}$, which implies $T_{\bm a} H_{\bm 0} T_{\bm a}^{-1} = H_0$, where $T_{\bm a}$ are the unitary operators of the lattice translations $T_{\bm a}|\bm n,\alpha \rangle = |\bm n + \bm a, \alpha \rangle $, $\bm a \in \mathbb Z^d$. The presence of a uniform magnetic field $\bm B$ immediately breaks the translational symmetry. It is true that, if all the fluxes: 
\begin{equation}
\Phi_{ij} = \bm B \cdot (\bm e_i \times \bm e_j), 
\end{equation}
 through the walls of the unit cell are rational numbers in units of the flux quantum $\phi_0 = h/\rm e$, then the periodicity can be restored by properly redefining the unit cell, but the probability of fine tuning a magnetic field to obey such a rational flux condition is zero. The effect of $\bm B$ is incorporated in the lattice models via the Peierls substitution \cite{Pei}:
\begin{equation}
t_{{\bm n}-{\bm m}}^{\alpha \beta} \rightarrow e^{ \I \, \bm n \wedge \bm m} t_{{\bm n}-{\bm m}}^{\alpha \beta}, \quad
\bm n \wedge \bm m = \frac{\pi}{\phi_0} \sum_{i,j=1}^d \Phi_{ij} n_i m_j.
\end{equation} 
Hence, the lattice Hamiltonians of periodic solids subjected to uniform magnetic fields take the generic form:
\begin{equation}
H_{\bm 0}(\bm B) = \sum_{{\bm n},\alpha; \bm m, \beta} e^{\I \, \bm n \wedge \bm m} \ t_{{\bm n}-{\bm m}}^{\alpha \beta} \; |{\bm n},\alpha \rangle  \langle {\bm m},\beta |.
\end{equation}
The Hamiltonians remain invariant with respect to the magnetic translations:
\begin{equation}
U_{\bm a} |\bm n,\alpha \rangle = e^{\I \, \bm n \wedge \bm a} |\bm n + \bm a, \alpha \rangle,
\end{equation}
which generate a projective representation of the additive group $\mathbb Z^d$:
\begin{equation}
U_{\bm a} U_{\bm b} = e^{\I \, \bm a \wedge \bm b} U_{\bm a + \bm b}, \quad \bm a, \bm b \in \mathbb Z^d.
\end{equation}
In real solids, there are other translation-breaking factors, such as the displacements of the atoms due to thermal motion or due to the inherent defects induced by the fabrication process. In such real-world conditions, the lattice Hamiltonians take the form:
\begin{eqnarray}\label{GModel}
H(\bm B)= \sum_{{\bm n},\alpha; \bm m, \beta} e^{\I \, \bm n \wedge \bm m} \ t_{{\bm n},{\bm m}}^{\alpha \beta} \; |{\bm n},\alpha \rangle  \langle {\bm m},\beta |,
\end{eqnarray}
where one should notice that the hopping amplitudes in Eq.~\ref{GModel} no longer depend on the difference $\bm n - \bm m$. The translation invariance is lost, even with respect to the magnetic translations.

\vspace{0.2cm}

Among the aperiodic systems, there is the special class of the homogeneous aperiodic systems mentioned at the beginning, which are translation invariant at a macroscopic scale but not necessarily at a microscopic one. The defining physical characteristics of these systems are the well defined mesoscopic transport coefficients, despite of the aperiodic and sometime disordered character of the samples. Examples are the quasicrystals, the amorphous solids and the disordered crystals. The difference between the last two is that the crystalline order is still present and experimentally detectable for the latter. In the present work, we are dealing exclusively with the homogeneous disordered crystals, which can be formalized as follows. First, let us assume that, due to the persistence of the crystalline order, we can formulate the models on the same Hilbert space $\mathcal H$ even though this might require a large number of orbitals per unit cell.  The Hamiltonians $H_{\bm \omega}(\bm B)$ are defined by the disordered configurations of the atoms, which are quantified by points $\bm \omega$ in a disorder configuration space $\Omega$. By adjoining all the translates of $H_{\bm \omega}(\bm B)$, if necessary, the family $\{H_{\bm \omega} (\bm B)\}_{\bm \omega \in \Omega}$ can be assumed invariant with respect to the magnetic lattice translations, that is, as an un-ordered family, $\{U_{\bm a} H_{\bm \omega}(\bm B) U_{\bm a}^{-1}\}_{\bm \omega \in \Omega}$ is identical to $\{H_{\bm \omega}(\bm B) \}_{\bm \omega \in \Omega}$. This has the following simple consequences:

\begin{enumerate}

\item Given any $H_{\bm \omega}(\bm B)$ from the family, then $U_{\bm a} H_{\bm \omega}(\bm B) U_{\bm a}^{-1}=H_{\bm \omega'}$ for some $\bm \omega' \in \Omega$. 

\item The pairs $(\bm \omega,\bm \omega')$ appearing above defines a bijective map 
\begin{equation}
\mathfrak{t}_{\bm a}:\Omega \rightarrow \Omega, \quad \mathfrak{t}_{\bm a}\bm \omega = \bm \omega', \quad \mathfrak t_{\bm a}^{-1} = \mathfrak t_{- \bm a}.
\end{equation}

\item The collection of maps $\{\mathfrak{t}_{\bm a} \}_{\bm a \in \mathbb Z^d}$ define an action of the additive $\mathbb Z^d$ group on $\Omega$, as one can readily verify that $\mathfrak{t}_{\bm a} \circ \mathfrak{t}_{\bm b} = \mathfrak{t}_{\bm a+\bm b}$.
\item The disordered Hamiltonians are covariant with respect to the magnetic lattice translations:
\begin{equation}
U_{\bm a}H_{\bm \omega}(\bm B) U_{\bm a}^{-1} = H_{\mathfrak t_{\bm a} \bm \omega}(\bm B), \ \ \mbox{for all} \ \bm a \in \mathbb Z^d.
\end{equation}

\item Furthermore, $\Omega$ can be topologized by identifying $\Omega$ with $\{H_{\bm \omega}(\bm B) \}_{\bm \omega \in \Omega}$, and seeing the latter as a subset of $\mathcal B(\mathcal H)$, the space of linear operators over $\mathcal H$ endowed with the strong topology. 

\end{enumerate}

Clearly, all the above can be applied to any aperiodic crystal, since the set $\Omega$ can be constructed from one single representative Hamiltonian. What makes $\{H_{\bm \omega}\}_{\bm \omega \in \Omega}$ into a homogeneous system is the fact that the closure of the set $\Omega$, when topologized as above, is compact. In such conditions, by replacing $\Omega$ with its closure, $(\Omega,\mathfrak t, \mathbb Z^d)$ becomes a classical dynamical system and, as such, it accepts at least one ergodic and invariant probability measure. If the system is translational invariant at the macroscopic scale, the physical probability measure must be one of them and will be denoted by $d \mathbb P(\bm \omega)$. A direct application of Birkhoff's ergodic theorem \cite{Bir}, which states that, $\mathbb P$-almost sure:
\begin{equation}\label{Birkhoff}
\lim_{V \rightarrow \mathbb Z^d} \frac{1}{|\rm V|}\sum_{\bm a \in V} f(\mathfrak{t}_{\bm a} \bm \omega)=\int_\Omega d \mathbb P(\bm \omega) f(\bm \omega),
\end{equation}
with $V$ a cube from $\mathbb Z^d$ and $|V|$ its cardinal, reveals the following self-averaging principle:
\begin{align}\label{TraceCell}
\lim_{V \rightarrow \mathbb Z^d}\frac{1}{|\rm V|} \sum\limits_{{\bm n} \in \rm V}\sum_{\alpha=1}^N \langle {\bm n},\alpha| F_{\bm \omega} | \bm n,\alpha \rangle
& =\lim_{V \rightarrow \mathbb Z^d} \frac{1}{|\rm V|} \sum\limits_{{\bm n} \in \mathrm{V}} \sum_{\alpha=1}^N \langle \bm 0,\alpha| F_{t_{\bm n}^{-1}\bm \omega} | \bm 0,\alpha \rangle  \\
     &=  \int_\Omega d \mathbb P(\bm \omega) \; \sum_{\alpha=1}^N \langle \bm 0,\alpha | F_{\bm \omega} | \bm 0, \alpha  \rangle,  \nonumber
\end{align}
for any covariant family of operators $F_{\bm \omega}$ affiliated with the homogeneous system. This shows explicitly why the intensive thermodynamic variables, in particular, the linear transport coefficients, do not fluctuate from one disorder configuration to another. This in turns explains why the macroscopic measurements on the mesoscopic samples mentioned at the beginning of this Section all return the same values, despite of the differences at the microscopic level.

\vspace{0.2cm}

The righthand side of Eq.~\eqref{TraceCell} is nothing but the trace per volume and will be denote in the following by $\mathcal T(\ldots )$. As opposed to the standard trace ${\rm Tr}$ on $\mathcal B(\mathcal H)$, $\mathcal T$ is normalized in the sense that $\mathcal T( I )=1$ while ${\rm Tr}(I) = \infty$. In general, the trace per volume can be computed using its very definition or as a disordered average, and both methods have their advantages. For example, the defining formula is a better choice in the numerical applications because of the self-averaging property which reduces the statistical fluctuations of the output values. The second method is useful in the analytic analysis because the integral over $\Omega$ can be manipulated like any other integral and, for example, changes of variables $\bm \omega \rightarrow \mathfrak t_{\bm a} \bm \omega$ show up quite often in the calculations,  in which case the invariance of $\mathbb P(d \bm \omega)$ w.r.t. translations proves to be a very useful property.  

\vspace{0.2cm}

We conclude this section with an explicit example of a homogeneous disordered lattice model:
\begin{equation}\label{EModel}
H_{\bm \omega}(\bm B)= \sum_{{\bm n},\alpha; \bm m, \beta} e^{2\pi \I \, \bm n \wedge \bm m} \big ( t_{{\bm n}-{\bm m}}^{\alpha \beta} +W \omega^{\alpha \beta}_{{\bm n},{\bm m}} \big )|{\bm n},\alpha \rangle  \langle {\bm m},\beta | ,
\end{equation}
where $\omega^{\alpha \beta}_{{\bm n},{\bm m}}$ are independent random entries drawn uniformly from the interval $\big [-\frac{1}{2},\frac{1}{2}\big]$. The collection of all random variables $\bm \omega=\{ \omega^{\alpha \beta}_{{\bm n},{\bm m}}\}$ can be viewed as a point in an infinite dimensional configuration space $\Omega$, which is just the infinite product of intervals $[-\frac{1}{2},\frac{1}{2}]$. This is a compact Tychonov space which can be equipped with the product probability measure: 
\begin{equation}
d \mathbb P(\bm \omega)=\prod_{\bm n,\alpha;\bm m, \beta} d\omega^{\alpha \beta}_{{\bm n},{\bm m}}.
\end{equation} 
There is a natural action of the lattice translations on $\Omega$:
\begin{equation}
(\mathfrak{t}_{\bm a} \bm \omega)^{\alpha \beta}_{{\bm n},{\bm m}}=\omega^{\alpha \beta}_{{\bm n}-{\bm a},{\bm m}-{\bm a}}, ~ \ a\in \mathbb{Z}^{D},
\end{equation}
which acts ergodically and leaves $d \mathbb P(\bm \omega)$ invariant, hence Eq.~\ref{Birkhoff} applies. It is easy to check that the Hamiltonian has indeed the covariant property:
\begin{equation}
U_{\bm a} H_{\bm \omega} U_{\bm a}^{-1} = H_{\mathfrak t_{\bm a} \bm \omega}.
\end{equation}
The disordered Hofstadter model is a particular case of the general model presented above. It is defined in dimension $d=2$ and it has only one orbital per unit cell, hence the index $\alpha$ can be dropped out:
\begin{equation}\label{HModel}
H_{\bm \omega}(\bm B)= \sum_{|\bm n-\bm m|=1} e^{\I \, \bm n \wedge \bm m} |{\bm n} \rangle  \langle {\bm m} | + W \sum_{\bm n}  \omega_{\bm n} |{\bm n}\rangle  \langle {\bm n} |.
\end{equation}
The wedge product can be written more explicitly
\begin{equation}
\bm n \wedge \bm m = \pi \frac{\Phi_{12}}{\phi_0}(n_1 m_2 - m_1 n_2).
\end{equation}
This model will be used in our numerical analysis. 

\section{The current-current correlation measure}

Assume a homogenous disordered crystal $H_{\bm \omega}(\bm B)$ as defined in the previous Section. Then the current-current correlation measure is related to the following correlation function, involving twice the current operator:
\begin{equation}
 \mathcal T \Big ( J_i(\bm B) \, \Phi \big ( H_{\bm \omega}(\bm B) \big ) \, J_j (\bm B) \, \Phi' \big ( H_{\bm \omega}(\bm B) \big ) \big ),
 \end{equation}
where $\Phi$ and $\Phi'$ are arbitrary first order differentiable functions defined on the real axis. We recall that the current operator is given by:
\begin{equation}\label{Current}
\bm J(\bm B) = \frac{{\rm e}}{\I \hbar} \big [ \bm X,H_{\bm \omega}(\bm B)\big ],
\end{equation}
where $\bm X$ is the position operator $\bm X|\bm n,\alpha \rangle = \bm n |\bm n,\alpha \rangle$. Note that, in our particular setting where the bond disorder is absent in $H_{\bm \omega}(\bm B)$, the current operator involves only the non-random part of the Hamiltonian, hence it is independent of the disorder configuration but it still depends on the magnetic field. Now, a rigorous result \cite{KP} states that there exists the Radon measures ${\rm d} m_{ij}(E,E')$ on $\mathbb R \times \mathbb R$ such that:
\begin{equation}\label{CCCMeasure}
  \mathcal T \Big ( J_i(\bm B) \, \Phi \big ( H_{\bm \omega}(\bm B) \big ) \, J_j(\bm B) \, \Phi' \big ( H_{\bm \omega}(\bm B) \big ) \big ) = \frac{{\rm e}^2}{\hbar^2} \int\limits_{\mathbb R \times \mathbb R}  \Phi(E) \, \Phi'(E') \, {\rm d} m_{ij} (E,E').
 \end{equation}
This equality defines the current-current correlation measure, consisting of the matrix of measures ${\rm d} m_{ij} (E,E')$. The support of the measure is $\Sigma \times \Sigma$, where $\Sigma$ is the $\mathbb P$-almost sure non-random spectrum of $H_{\bm \omega}(\bm B)$. We will denote its isotropic part by: 
\begin{equation}
{\rm d} m(E,E')=\frac{1}{d}\sum_{j=1}^d {\rm d} m_{jj}(E,E').
\end{equation} 
As we already pointed out in our introduction and will be further seen below, $d m_{ij} (E,E')$ plays a central role in the theory of random Schr\"oedinger operators. 

\vspace{0.2cm}

The transport theory of homogeneous aperiodic solids in the presence of dissipation is now well established and detailed accounts of it can be found in many references \cite{BES,SBB1,SBB2,AFG}. See also the review in Ref.~\cite{Bel1} or Ref~\cite{Pro2} for a computational perspective. There are of course many other references, notably, \cite{BGKS,Kun,Nak}, but there the dissipative processes, which determine the critical behavior, are ignored. The Kubo-formula with dissipation, in its full generality reads (\cite{SBB1}, Eq.~28):
\begin{equation}\label{Kubo1}
\sigma_{ij}(T,E_F)=\I \frac{\rm e}{\hbar}\mathcal{T}\Big (J_i(\bm B)\big (\Gamma+\mathcal{L}_H \big )^{-1} \big [X_j,\Phi_{\beta}(H_{\bm \omega}(\bm B)-E_F)\big ] \Big ).
\end{equation}
Here, $\mathcal{L}_H$ represents the Liouvillian super-operator acting on operators as $\mathcal{L}_H(A) = \I [A,H]$, $\Gamma$ is the dissipation super-operator which has a temperature dependence, in general, and $\Phi_{\beta}$ is the Fermi-Dirac distribution $\Phi_{\beta}(x)=\frac{1}{1+e^{\beta x}}$, 
depending parametrically on the temperature via $\beta = 1/kT$. Also, $E_F$ in Eq.~\eqref{Kubo1} represents the Fermi level. Various models for the dissipation super-operator $\Gamma$ and the physical regimes where they are expected to apply are discussed in Refs.~\cite{SBB2,SpB,Bel1,ABS}. Numerical applications of the Kubo-formula with dissipation can be found in Refs.~\cite{SP,Pro2,XP1,XP2}, which focused on the critical behavior of the transport coefficients of disorder topological insulators.

\vspace{0.2cm}

In this work, we will restrict to the so called relaxation time approximation, which assumes $\Gamma$ proportional to the identity $\Gamma = \hbar/\tau$, with $\tau$ a $c$-number commonly referred to as the relaxation time \cite{SBB1}. In this case, the isotropic direct conductivity of a homogenous systems was shown \cite{SBB2} to accept the following formula in terms of the current-current correlation measure (\cite{SBB1}, Eq.~29):
\begin{equation}\label{Kubo2}
\sigma(\beta,E_F,\Gamma) = \frac{\rm e^2}{h} \int\limits_{\mathbb R \times \mathbb R} \frac{\Phi_{\beta}(E'-E_F) - \Phi_{\beta}(E-E_F)}{E-E'} \frac{4\pi \Gamma}{\Gamma^2 + (E-E')^2} {\rm d} m (E,E').
\end{equation}

The Anderson localization length also accepts a formula in term of the current-current correlation measure. Let us recall first the so called $\Delta$-localization length $\Lambda(\Delta)$ \cite{BES}, with $\Delta$ an interval of $\mathbb R$ centered at the Fermi level:
\begin{equation}
\Lambda^2(\Delta) = \mathcal T \Big ( \big | [P_\Delta, \bm X] \big |^2 \Big ).
\end{equation}
Here, $P_\Delta$ the spectral projector of $H_{\bm \omega}(\bm B)$ onto $\Delta$: $$
P_\Delta = \chi \big (H_{\bm \omega}(\bm B) \in \Delta \big ).
$$ 
Then \cite{BES}:
\begin{equation}
\Lambda^2(\Delta) = \int_{\Delta \times \mathbb R} \frac{{\rm d} m (E,E')}{(E-E')^2}.
\end{equation}
It is a well established fact \cite{BES,SBB2} that, whenever there exists a finite interval $\Delta$ centered at $E_F$ and such that $\Lambda(\Delta) < \infty$, the direct conductivities $\sigma_{ii}$ vanish in the limit $T \searrow 0$. The Anderson localization length, defined strictly at the Fermi level, can be expressed as:
\begin{equation}
\Lambda^2(E_F) = \lim_{|\Delta| \rightarrow 0} \frac{\Lambda^2(\Delta)}{|\Delta|} = \lim_{|\Delta| \rightarrow 0} \int_{\Delta \times \mathbb R} \frac{{\rm d} m (E,E')}{(E-E')^2},
\end{equation}
where by $|\Delta|$ is meant the length of the interval. 

\begin{figure}
\center
{\includegraphics[width=0.8\columnwidth]{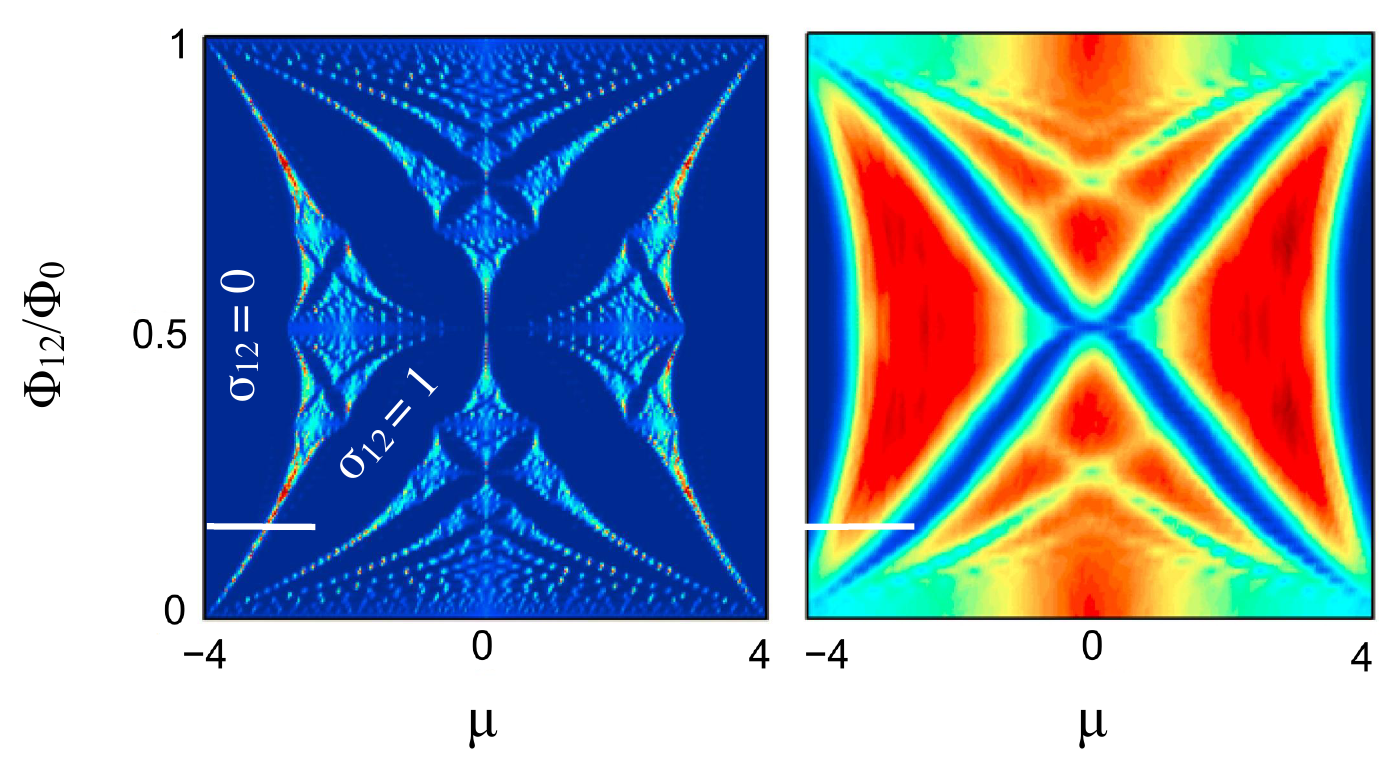}}
\caption{The local density of states of the model \eqref{EModel}, as computed from $\frac{1}{\pi}{\rm Im}\; (H_{\bm \omega}(\bm B) + 0.01\I)^{-1}(\bm n,\bm n)$, plotted as an intensity map in the plane of Fermi level $E_F$ and magnetic flux per unit cell $\Phi_{12}$. The left/right panels correspond to $W=0$ and $W=3$, respectively. This data is rather used for identifying the broad features of the spectrum and not for understanding the local density of states itself. The horizontal white lines show the range of energies and the value of the magnetic flux considered in the numerical simulations.}
\label{DOS}
\end{figure}

\section{Behavior near a critical point}

Let us use the concrete disordered Hofstadter model of Eq.~\ref{HModel} to set the stage. The local density of states of the model is shown in Fig.~\ref{DOS} as an intensity map in the plane $(E_F,\Phi_{12})$, for both the clean and disordered cases. In the clean case, one can observe the presence of several prominent spectral gaps and the values of the Hall conductivity $\sigma_{12}$ at $T=0$ are shown in Fig.~\ref{DOS} for two of these gaps. It is well known that $\sigma_{12}$ is proportional to the Chern number \cite{BES}, which is a topological invariant whose quantized value can change only if the Fermi level crosses a spectral region were the direct conductivity is strictly positive. This is indeed what must happen as one crosses the region separating the prominent spectral gaps. In fact, the Chern number changes its value an infinite number of times because of the fractal structure of the Hofstadter butterfly \cite{AO,AOS,AKY,AEG}. In the disordered case, shown in the right panel of Fig.~\ref{DOS}, the strength $W$ was chosen so that the prominent spectral gaps do not close completely so that we can be sure that at least two different quantum Hall phases survive the disorder. Yet $W$ was chosen large enough so that the fractal structure was washed away and, as the Fermi level transits along the horizontal line in Fig.~\ref{DOS}, there is only one localization-delocalization transition. Experimentally, the precise location of the quantum phase transition between two IQHE phases  is detected by tracing the conductivity tensor as function of Fermi energy (or rather the electron density) for decreasing values of temperature. Once the temperature is low enough, these traces display the single-parameter scaling paradigm where all the curves fall on top of each other after a rescaling of the energy axis:
\begin{equation}\label{scaling}
E_F \rightarrow E_c + (E_F -E_c) \left ( \frac{\beta}{\beta_0} \right )^\kappa,
\end{equation}
with $\beta_0$ a reference temperature. As we already mentioned in our introduction, this is sharply reproduced by most experimental data. When this scaling law applies with great accuracy, the system is said to have entered the quantum critical regime. Additionally, the Anderson localization length diverges at the critical point as in Eq.~\ref{EScaling}, 
with an exponent $\nu$ which relates to $\kappa$ as in Eq.~\ref{Kappa}.

\vspace{0.2cm}

Deriving an asymptotic behavior of the current-current correlation measure which explains such critical behavior is the main goal of the present Section. Since the only input for our analysis is the invariance of the direct conductivities against the scaling law \eqref{scaling} and the diverging behavior of Anderson's localization length, the results of this Section are not bound to the IQHE context. 

\vspace{0.2cm}

As we already mentioned in the introduction, our main assumption is that the current-current correlation measure is continuous w.r.t. the Lebesgue measure:
\begin{equation}
{\rm d} m(E,E') = f(E,E') \, {\rm d}E {\rm d}E',
\end{equation}
with $f$ continuous in both variables. The latter is referred to as the current-current correlation function. From the beginning, we will render the energies $E$, $E'$ and $E_F$ from the critical value. Hence, from now on, the critical point occurs at the origin. It will be convenient to perform the changes of variables:
\begin{equation}
x=\tfrac{1}{2}(E+E'), \ \ y = \tfrac{1}{2}(E-E'),
\end{equation}
hence:
\begin{equation}
{\rm d} m(E,E') = f(E,E') \, {\rm d}E {\rm d}E' = \tfrac{1}{2} f(x,y) \, {\rm d}x {\rm d}y.
\end{equation}
Using the cyclic property of the trace per volume $\mathcal T$, it follows automatically from \eqref{CCCMeasure} that $f(E,E')=f(E',E)$, which then implies $f(x,y) = f(x,-y)$. As such, we will write 
\begin{equation}
{\rm d} m(E,E')= \frac{1}{2}f(x,|y|) \, {\rm d}x {\rm d}y,
\end{equation} 
from now on. Then:
\begin{equation}
\sigma(\beta,E_F,\Gamma) = \frac{\rm e^2}{h} \int\limits_{\mathbb R \times \mathbb R} \frac{\Phi_{\beta}(x-E_F+y) - \Phi_{\beta}(x-E_F-y)}{2y} \frac{2\pi \Gamma}{\Gamma^2 + y^2}f(x,|y|) {\rm d}x {\rm d}y.
\end{equation}
Recall that $\Gamma \searrow 0$ as $T \searrow 0$, hence both first two factors inside the integral are approximates of the Dirac-delta distribution. However, the convergence of the first factor is much faster and we will assume that we are in a regime where:
\begin{equation}
\frac{\Phi_{\beta}(x-E_F+y) - \Phi_{\beta}(x-E_F-y)}{2y} \approx - \beta \Phi'_\beta(x-E_F) \rightarrow \delta(x-E_F),
\end{equation} 
while the second factor is far from such limit. Hence:
\begin{equation}\label{Step1}
\sigma(\beta,E_F,\Gamma) = \frac{\rm e^2}{h} \int_{-\infty}^\infty \frac{2\pi \Gamma}{\Gamma^2 + y^2}f(E_F,|y|) {\rm d}y = \frac{\rm e^2}{h} \int_0^\infty \frac{4\pi}{1+ y^2}f (E_F,\Gamma y) {\rm d}y.
\end{equation}
Note that $\sigma$ is now a function of only $E_F$ and $\Gamma$, hence we will write $\sigma(E_F,\Gamma)$ from now on. The scaling law of the conductivity observed in the quantum critical regime can be interpreted as an invariance relative to the transformation:
\begin{equation}
\sigma(\lambda^\kappa E_F, \lambda^{p} \Gamma) = \sigma(E_F,\Gamma), 
\end{equation}
where $\lambda$ is a scaling factor with the range in $[0,\infty)$. When applied to Eq.~\ref{Step1}, we obtain:
\begin{equation}\label{Step2}
\int_0^\infty \frac{\pi}{1+ y^2}\big [f (\lambda^\kappa E_F,\lambda^p\Gamma y ) -f(E_F,\Gamma y) \big ] {\rm d}y = 0.
\end{equation}
Of course, there are many solutions to this equation, but the most obvious one is:
\begin{equation}
f(\lambda^\kappa x,\lambda^p y) = f(x,y).
\end{equation}
If we adopt this solution, we can recast the current-current correlation measure in terms of a single-variable function:
\begin{equation}\label{DGFunction}
\boxed{
f(x,y) = g\left( \frac{x}{|y|^{\kappa/p}}\right ),
}
\end{equation}
where it is assumed that $g$ has limits at $0$ and $\infty$. In summary, we propose the following formula:
\begin{equation}
\boxed{
\sigma(E_F,\Gamma)= \frac{\rm e^2}{h} \int_0^\infty \frac{4\pi}{1+ y^2} g\left(\frac{E_F}{(\Gamma y)^{\kappa/p}} \right ) {\rm d}y,
}
\end{equation}
which, as we have shown above, captures the essential behavior of the isotropic transport coefficient in the quantum critical regime. Furthermore, the asymptotic behavior of $g$, at both the origin and at infinity, can be settled using the fact that the transition is between two insulating phases and that the isotropic conductivity settles at a finite value exactly at the critical point as temperature is taken to zero. This can be rephrased as:
\begin{equation}
\lim_{\Gamma \searrow 0} \sigma(E_F,\Gamma) = \left \{
\begin{aligned}
& 0, \quad \quad \quad \quad E_F \neq 0 , \\
& \sigma_c < \infty, \quad \ \ E_F=0.
\end{aligned}
\right.
\end{equation}
From the first limiting behavior, we conclude that
\begin{equation}\label{Asympt1}
\lim_{t \rightarrow \infty} g(t) =0,
\end{equation}
while from the second one:
\begin{equation}\label{Asympt2}
\lim_{t \rightarrow 0} g(t) =\frac{\sigma_c}{2\pi^2 {\rm e^2}/h}.
\end{equation}

We now turn our attention to the scaling law of the localization length near the critical point. Of course, the temperature is out of the picture and the focus is on the diverging behavior of $\Lambda(E_F)$ as $E_F \rightarrow 0$. We have:
\begin{equation}\label{Lambda}
\Lambda^2(E_F) = \int_{- \infty}^\infty \frac{f(E_F,E) {\rm d}E}{(E_F-E)^2} =\int_{- \infty}^\infty \frac{1}{(E_F-E)^2} \, g \left ( \frac{E_F+E}{|E_F-E|^{\kappa/p}} \right ) {\rm d}E,
\end{equation}
or, after the change of variable $y=E_F-E$:
\begin{equation}\label{Inter1}
\Lambda^2(E_F) =  2\int_0^\infty y^{-2} g \left ( \frac{2E_F}{y^{\kappa/p}} + y^{1-\kappa/p} \right ) {\rm d}y.
\end{equation}
We assume that $g$ is continuous and that $\kappa/p <1$, which we already know to be the case for IQHE \cite{LangPRB2007yg}. If we set $E_F=0$, then, based on the asymptotic behavior of $g$ from Eq.~\ref{Asympt2}, one can readily see that the integral is divergent because of the behavior of the integrand for $y$ close to the origin. For $E_F \neq 0$, the integral is finite, provided the decay of $g$ to zero as the argument goes to infinity is fast enough (which we need to assume anyway to ensure $\sigma < \infty$). One can also see that the integral is convergent for $y$ away from the origin, for all values of $E_F$ including $E_F=0$. The conclusion is that the divergence of $\Lambda(E_F)$ as $E_F \rightarrow 0$ originates from the integration of $y$ near the origin. In this region of integration, we can drop the second term in the argument of $g$ in Eq.~\eqref{Inter1}, and write:
 \begin{equation}
\Lambda^2(E_F) \sim 2 \int_0^\infty y^{-2} g \left ( \frac{2E_F}{y^{\kappa/p}} \right ) {\rm d}y = \frac{2}{(2 E_F)^{p/\kappa}} \int_0^\infty g(y^{\kappa/p}){\rm d}y,
\end{equation}
where in the last equality we performed an appropriate change of variable. The remaining integral is just a finite $c$-number because the integral is convergent and, as a consequence, our solution Eq.~\eqref{DGFunction} reproduces Eq.~\eqref{EScaling}, provided $p/\kappa = 2\nu$. The latter is precisely the Thouless rule of Eq.~\eqref{Kappa}.

\vspace{0.2cm}

We end this Section with several predictions:

\begin{enumerate}[{\rm (i)}]

\item From the limit listed in Eq.~\eqref{Asympt1}, it follows that $f(E,E')$ is exactly zero along the diagonal $E=E'$, except at the critical point where the exact value is indeterminate.

\item From the previous item, if the critical point $(E_c,E_c)$ is approached along the diagonal $E=E'$, then the limit value of $f(E,E')$ is zero. But if the critical point $(E_c,E_c)$ is approached from any other direction, the limit value of $f(E,E')$ is $\frac{\sigma_c}{2\pi^2 {\rm e^2}/h}$, which follows from the limit listed in Eq.~\ref{Asympt2}.

\item The level sets of $f(E,E')=g(t)$ near the critical point are well described by the equation
\begin{equation}\label{Match}
E+E' = t (E-E')^\frac{1}{2\nu}.
\end{equation}
This observation can also be used to map the function $g$ of Eq.~\ref{DGFunction}.
\end{enumerate}

\section{Numerical analysis}

The primary goal of this section is to demonstrate that, in principle, we do have access to the exact current-current correlation measure, provided the assumptions made in the previous Section hold. Let us point out that the methodology and the numerical algorithm reported below are not specific to the present context and they can be implemented to any lattice model. Another goal is to demonstrate qualitatively and semi-quantitatively that the predictions made in the previous Section for the Hofstadter model are confirmed by the numerical results. Note that in this case it is known from experiment \cite{VisserJPCS2006cu} as well as simulations \cite{SP} that $\sigma_c = \frac{1}{2}\frac{{\rm e}^e}{h}$, hence the expected value of the current-current correlation function at the critical point is $f(E_c,E_c) = 1/4 \pi ^2$.

\vspace{0.2cm}

Let us first lay down the basic principles of our numerical algorithms. Consider an approximation of the Dirac-delta distribution:
\begin{equation}\label{DiracDelta}
\delta_\epsilon(t) = \frac{1}{\epsilon} \phi\left(\frac{t}{\epsilon}\right ) \overset{\epsilon \rightarrow 0}{\longrightarrow}\delta(t),
\end{equation}
where $\phi$ is a smooth real-valued function which integrates to 1. With our continuity assumption on $f$, it is immediate to see that:
\begin{equation}\label{FEpsilon}
f_\epsilon(E,E')= \int\limits_{\mathbb R \times \mathbb R}  \delta_\epsilon(t-E) \delta_\epsilon(t-E') \, {\rm d} m (t,t')
\end{equation}
is a point-wise approximate of $f(E,E')$: 
\begin{equation}\label{Limit}
\lim_{\epsilon \rightarrow 0} f_\epsilon(E,E') = f(E,E'), \quad E, E' \in \mathbb R.
\end{equation}
The point here is that the distribution $f_\epsilon$ can also be computed from:
\begin{equation}\label{Principle}
f_\epsilon(E,E') =\frac{\hbar^2}{{\rm e}^2} \frac{1}{d}\sum_{i=1}^d \mathcal T \Big ( J_i(\bm B) \, \delta_\epsilon \big ( H_{\bm \omega}(\bm B)-E \big ) \, J_i(\bm B) \, \delta_\epsilon \big ( H_{\bm \omega}(\bm B)-E'  \big ) \Big ).
 \end{equation}
which is amenable on a computer, as explained below. The above expression, together with the assuring fact of Eq.~\ref{Limit}, represent the basis for our numerical simulations. In the following, we will refer to $f_\epsilon$ as the $\epsilon$-approximation of the current-current correlation function $f$.
 
 \begin{figure}
\center
{\includegraphics[width=1.00\columnwidth]{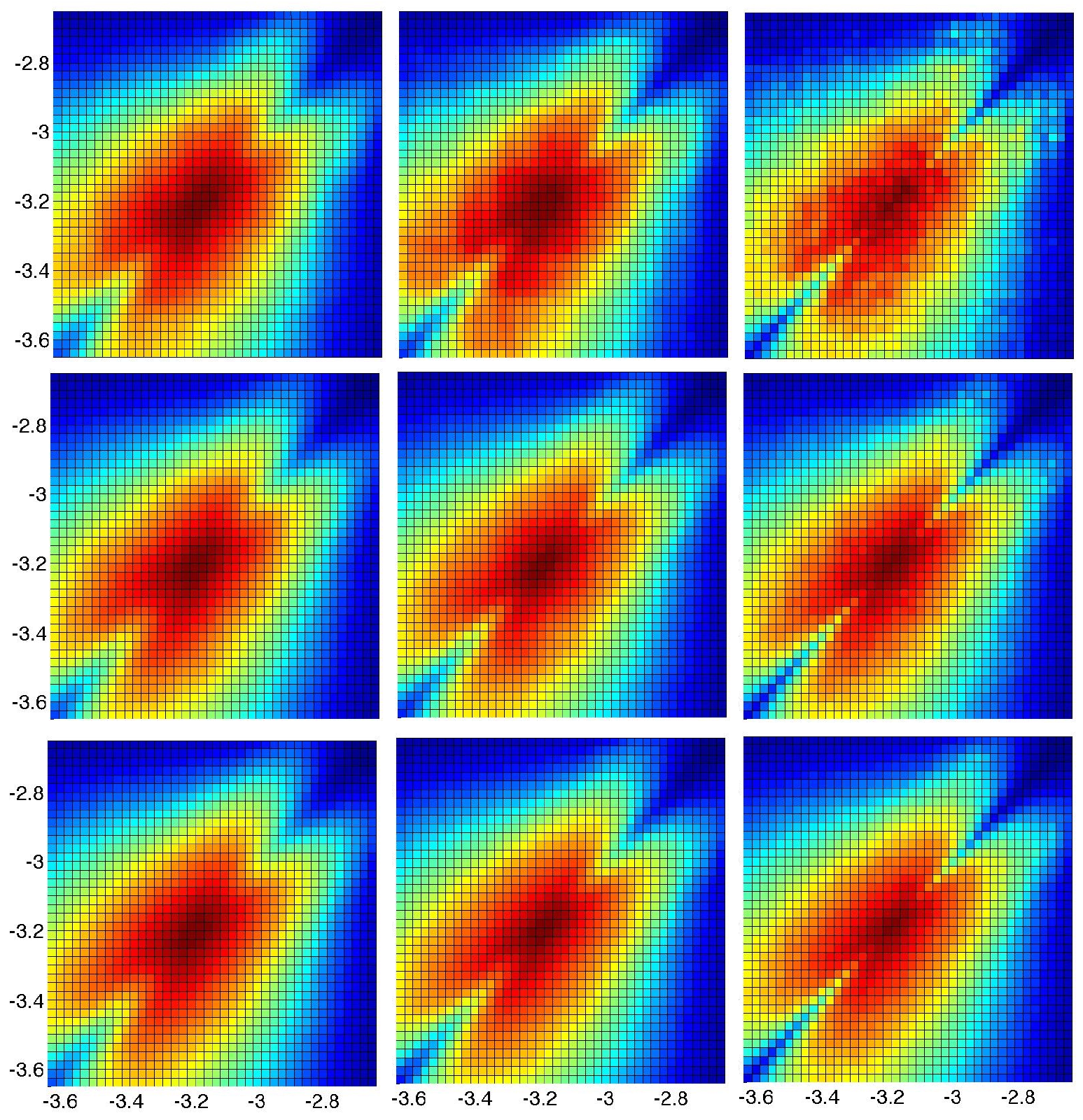}}
\caption{Convergence tests on the finite-size $\epsilon$-approximation $f_\epsilon^L(E,E')$ defined in Eq.~\eqref{FSPrinciple}, as computed numerically using a Gaussian profile $\phi(t) = \frac{1}{\sqrt{2\pi}}e^{-t^2/2}$. The first, second and third rows correspond to different lattice sizes: $40 \times 40$ and $80 \times 80$ and $120 \times 120$, respectively. The first, second and third columns correspond to different $\epsilon$ values (see Eq.~\ref{DiracDelta}): $\epsilon=0.03$, $0.02$ and $0.01$, respectively. Hence, the most accurate representation of the true current-current correlation function $f(E,E')$ in this figure is provided by the right-lower panel.}
\label{ConvCheck1}
\end{figure}
 
\vspace{0.2cm} 
 
The righthand side of Eq.~\ref{Principle} was implemented on a computer and the details of the simulation are as follows. We performed an exact diagonalization of the disordered Hofstadter model Eq.~\ref{HModel} on a square lattice of finite $L \times L$ size, with periodic conditions imposed at the boundaries. The latter is compatible with magnetic translations only for a set of quantized magnetic fluxes $\Phi_{12} = \frac{k}{L} \Phi_0$. In all our simulations, we fixed this quantized value at $\Phi_{12} = 0.15 \, \Phi_0$, which corresponds to the horizontal line crossing the lowest Landau band in Fig.~\ref{DOS}. The energies $E$ and $E'$ were varied along this line. This particular setup was chosen so that we agree with the previous simulations performed in Ref.~\cite{SP}. The strength of the random potential was fixed at $W=3$. The random potential was updated 100 times and the results presented here are averages over these disorder configurations. Lastly, let us note that the current operator, whose expression is given in Eq.~\eqref{Current}, involves only the non-random part of the Hamiltonian. Furthermore, since the Hamiltonian is short-range and the magnetic flux is properly quantized, the finite-size current operator $\bm J^L$ and the finite-size Hamiltonian $H_{\bm \omega}^L(\bm B)$ are generated by simply imposing the periodic boundary conditions (and no truncations are necessary). The righthand side of Eq.~\eqref{Principle} on a finite size lattice then becomes:
\begin{equation}\label{FSPrinciple}
f_\epsilon^L(E,E') =\frac{\hbar^2}{{\rm e}^2} \frac{1}{d}\sum_{i=1}^d \mathcal T \Big ( J_i^L(\bm B) \, \delta_\epsilon \big ( H_{\bm \omega}^L(\bm B)-E \big ) \, J_i^L(\bm B) \, \delta_\epsilon \big ( H_{\bm \omega}^L(\bm B)-E'  \big ) \Big ).
 \end{equation}

The convergence of the simulations with respect to the finite size $L$ was studied theoretically in Ref.~\cite{Pro2}. Let $f_\epsilon^L(E,E')$ be the $\epsilon$-approximation computed on a finite lattice of size $L \times L$, as described above. Then the estimates derived in Ref.~\cite{Pro2} assures us of the following rapid convergence:
\begin{equation}\label{Conv}
\big |f_\epsilon^L(E,E') - f_\epsilon(E,E') \big | \leq A_\epsilon e^{- \gamma_\epsilon L},
\end{equation}
provided the function $\phi$ appearing in Eq.~\eqref{DiracDelta} is analytic in a strip around the spectrum $\Sigma$ of $H_{\bm \omega}(\bm B)$. These conditions are met if, for example, we choose the Gaussian $\phi(t) = \frac{1}{\sqrt{2\pi}}e^{-t^2/2}$ or the Lorentzian $\phi(t) = \frac{1}{\pi}\frac{1}{1+t^2}$ profiles. If this is the case, then the error bound in Eq.~\eqref{Conv} applies regardless of the localized or delocalized character of the spectrum but, of course, the numerical values of the coefficients depend on such details. They also depend on $\epsilon$ and the convergence with $L$ (at fixed $\epsilon$) is expected to slow down as $\epsilon \searrow 0$. The important remark is that, away from the singular points, we can achieve a good representation of $f(E,E')$ even with a finite $\epsilon$. How small we need to take $\epsilon$ depends on the profile of $f(E,E')$ and this is explored next. 

\vspace{0.2cm}

Fig.~\ref{ConvCheck1} reports a convergence test generated with a Gaussian profile. In these simulations, we computed $f_\epsilon^L(E,E')$ for $(E,E')$ on the grid $\mathcal G$ which is clearly visible in the plots, and the value of $\epsilon$ was reduced from $0.03$ to $0.01$ and the lattice-size was increased from $L=40$ to $L=120$. Let us make some qualitative remarks first. Although not a perfect representation of $f(E,E')$, we can already detect the position of the critical point and the numerical values at this point agree extremely well with the prediction (ii) of the previous Section. For example, the numerical value at the critical point for $\epsilon = 0.01$ and $L=120$ is $0.9981$ in units of $\frac{1}{4\pi^2}$. In fact, the overall shape is consistent qualitatively with the theoretical predictions. We can already see that the current-current correlation function is smooth away from the critical point, and even featureless away from the diagonal $E=E'$, but it displays abrupt features near critical point, especially near the diagonal where $f_\epsilon^L$ drops to lower values. This feature is consistent with the prediction (i) of the previous Section (and that of Ref.~\cite{CGH}), which says that $f(E,E)=0$ except at $E_c$, but to fully resolve the behavior near the diagonal it will be a very difficult numerical task. In Fig.~\ref{ConvCheck1}, one can clearly see how the structure near the diagonal becomes sharper as $\epsilon$ is decreased but the graph itself becomes more rugged, as expected. Increasing the system-size makes the graph smooth again. 

\begin{table}\label{Table}
\begin{center}

\begin{tabular}{|c|c|c|c|c|c|c|c|c|c|}
\hline
\ & $\epsilon=\epsilon' = 0.03$ & $\epsilon=\epsilon' = 0.02$ & $\epsilon =\epsilon'= 0.01$ 
\\\hline
$L=40, L'=80 \ \ $  & $1.1\times 10^{-2}$ & $1.5 \times 10^{-2}$ & $1.9 \times 10^{-2}$  
\\\hline 
$L=80, L'=120$ & $6.5 \times 10^{-3}$ & $6.8\times 10^{-3}$ & $1.1 \times 10^{-2}$
\\
\hline
\end{tabular}
\end{center}
\caption{The estimator $D$ evaluated on the data from Fig.~\ref{ConvCheck1}.}
\label{tab-class}
\end{table}

\vspace{0.2cm}

\begin{figure}
\center
{\includegraphics[width=0.9\columnwidth]{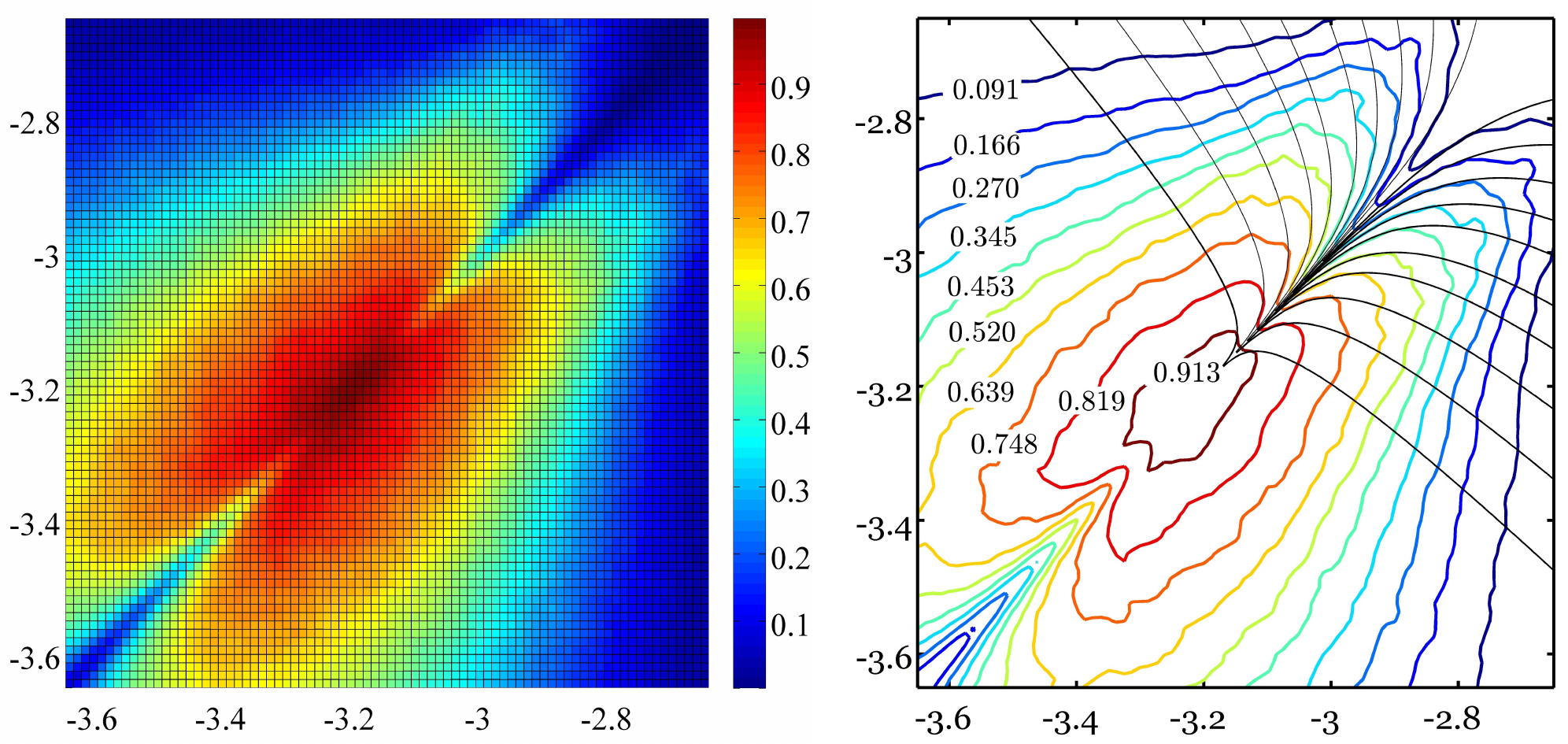}}
\caption{Left: An intensity plot of the current-current correlation distribution $f(E,E')$. Right: The level sets of $f(E,E')$ together with  matching curves described in Eq.~\ref{Match}. The computation was performed on a 120$\times$120 lattice and the data was averaged over 100 random configurations.}
\label{CCM}
\end{figure}

To quantify the convergence tests, we evaluated the following quantity:
\begin{equation}\label{Estimator}
D = \frac{4\pi^2}{|\mathcal G|}\sum_{(E,E') \in \mathcal G} \Big |f_\epsilon^L(E,E') - f_{\epsilon'}^{L'}(E,E') \Big |,
\end{equation}
which is an estimator of the (absolute) variations from one simulation to another, and we report the findings in Table~\ref{Table}. Based on these numbers, we expect that, for these tested values of $\epsilon$, the $\epsilon$-approximation $f_\epsilon(E,E')$ be converged w.r.t. the system-size to at least two digits of precision (in units of $\frac{1}{4\pi^2}$). To quantify how far is $f_\epsilon(E,E')$ from the true current-current correlation function, we evaluated the estimator at $L=120$ and found $D=9.5 \times 10^{-3}$ when decreasing $\epsilon$ from $0.03$ to $0.02$,  and $D= 7.2 \times 10^{-3}$ when decreasing $\epsilon$ from $0.02$ to $0.01$. We could infer from these numbers that, on average, the true current-current correlation function $f(E,E')$ was also resolved up to two digits of precision (in units of $\frac{1}{4\pi^2}$). 

\vspace{0.2cm}

\begin{figure}
\center
{\includegraphics[width=1.0\columnwidth]{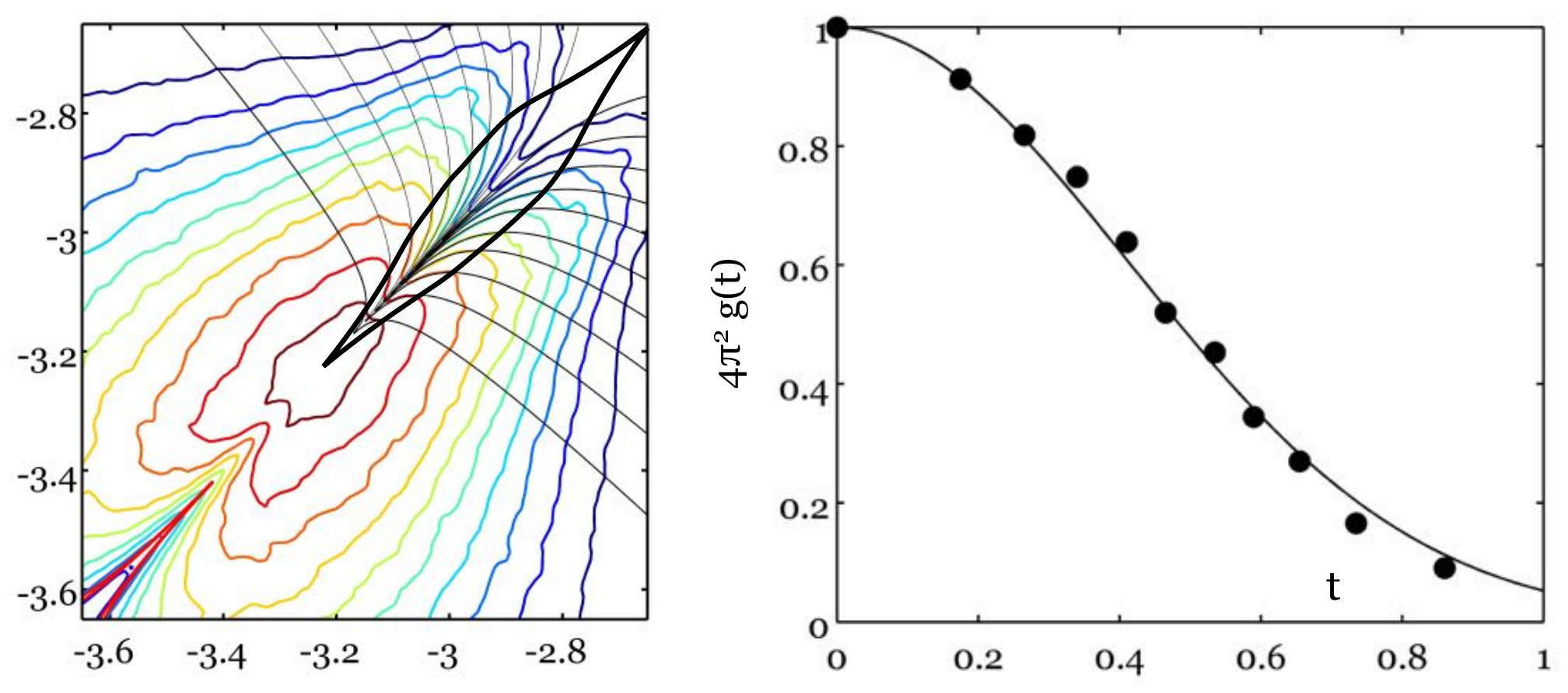}}
\caption{Left: The trace of the asymptotic region where the scaling invariance of the current-current correlation function occurs. Right: Plot of the 10 values of the function $g(t)$, as derived from the 10 contours from Fig.~\ref{CCM}, together with a Gaussian fit.}
\label{GFunction}
\end{figure}

We now analyze in more details the current-current correlation function. In Fig.~\ref{CCM}(a) we report the numerically computed  distribution $f_\epsilon^L(E,E')$, in units of $\frac{1}{4 \pi^2}$ and for $L=120$, $\epsilon=0.01$ but on a more refined grid than in Fig.~\ref{ConvCheck1} (which is again clearly visible from the graph). Based on our convergence tests, we believe that the data is an accurate representation of the exact $f(E,E')$. Panel (b) of Fig.~\ref{CCM} displays 10 level sets of $f(E,E')$, which will be used to test the prediction (iii) of the previous Section. For this we overlap in Fig.~\ref{CCM}(b) the matching contours generated with Eq.~\ref{Match}, $E =E' + t (E-E')^\frac{1}{2\nu}$. Here, we used $\nu = 2.58$ from the previous simulations \cite{SP} and only optimized the value of $t$ for each level set. Although the quality contours is somewhat low, the agreement between the numerical level sets and Eq.~\ref{Match} is surprisingly good in a region near the diagonal. Beyond this region, the two curves rapidly diverge from one another. This give us an estimate of the asymptotic region where the scaling invariance applies and this region is traced for the eye in the left panel of Fig.~\ref{GFunction}. Furthermore, by pairing the values of $t$ used to generate the matching contours in Fig.~\ref{CCM} with the values of the level sets, we can generate the profile of the function $g$ and this is shown in the right panel of Fig.~\ref{GFunction}. This profile is quite different from a Lorentzian or a Poisson profile but it is represented quite well by a Gaussian, as the fit shows.

\section{Conclusions}

In conclusion, using reasonable assumptions, we have shown that the critical behavior of the transport coefficients and of Anderson's localization length at a quantum phase transition are both a result of a particular asymptotic behavior of the current-current correlation function near the critical point. We have described a general numerical algorithm for generating $\epsilon$-approximations of the current-current correlation function of homogeneous solids. These approximations can provide a good representation of the correlation function if the latter is continuous. We have applied this algorithm to the disordered Hofstadter model and various convergence tests indicated that the current-current correlation function is indeed continuous except at the critical point. Furthermore, the numerical results reproduce the asymptotic behavior predicted theoretically in the first part of our work. 

 \section*{Acknowledgement} EP acknowledges financial support from the U.S. NSF grant DMR-1056168 and JB from the U.S. NSF grant DMS-1160962.

\end{document}